\documentclass[a4paper,USenglish,cleveref,autoref, authorcolumns]{lipics-v2021}

\usepackage{rotating}
\usepackage{adjustbox}
\usepackage{tikz}
\usepackage{filecontents}
\usepackage{xspace}
\usepackage{listings}
\usepackage{bm}
\usepackage{array}
\usepackage{longtable}
\usepackage{multirow}
\usepackage{minted}
\makeatletter
\@ifundefined{MintedVerbatim}
  {\DefineVerbatimEnvironment{MintedVerbatim}{Verbatim}{}}{}
\makeatother
\input{pygstyle}
\newsavebox{\pygbgbox}
\newenvironment{pygcolorbg}[1]
  {\def\pygbgcol{#1}%
   \par\noindent
   \begin{lrbox}{\pygbgbox}%
   \begin{minipage}{\dimexpr\linewidth-2\fboxsep\relax}}
  {\end{minipage}%
   \end{lrbox}%
   \colorbox{\pygbgcol}{\usebox{\pygbgbox}}}
\usepackage{makecell}
\usepackage{tcolorbox}
\tcbuselibrary{skins,breakable}
\usepackage{xcolor}
\definecolor{darkgrey}{rgb}{0.3,0.3,0.3}
\usepackage{pifont}
\usepackage{booktabs}
\usepackage{xurl}
\usepackage{amsmath,amsfonts}
\usepackage{textcomp}
\usepackage{fancyvrb}
\usepackage{graphicx}
\usepackage{subcaption}   
\usepackage{multicol}
\usepackage{algorithm}
\usepackage{algpseudocode}
\algrenewcommand\alglinenumber[1]{#1:}
\usepackage{changepage}
\usepackage{verbatim}
\usepackage[inkscapelatex=false]{svg}

\newcolumntype{?}{!{\vrule width 1pt}}

\newtcolorbox{mybox}[2][]{text width=0.95\linewidth,fontupper=\small,
fonttitle=\bfseries\sffamily\scriptsize, colbacktitle=darkgrey,enhanced,
attach boxed title to top left={yshift=-2mm,xshift=3mm},
boxed title style={sharp corners},top=4pt,bottom=2pt,left=2pt,right=2pt,
  title=#2,colback=white}

\newcommand{\ying}[1]{\textcolor{black}{{} #1}}

\newcommand{\lx}[1]{\textcolor{black}{{} #1}}

\newcommand{\oldx}[1]{\textcolor{black}{#1}}
\newcommand{\newx}[1]{\textcolor{black}{#1}}

\newcommand{\tool}{\textsc{CognixShield}\xspace}
\newcommand{\fo}{\textsc{GPT-4o}\xspace}
\newcommand{\gpto}{\textsc{GPT-o1}\xspace}

\newcommand{\claud}{\textsc{Claude-3.5-Sonnet}\xspace}
\newcommand{\gemini}{\textsc{Google-Gemini-2.0-Flash}\xspace}
\newcommand{\fb}{\textsc{Llama-3.1-405B}\xspace}

\newcommand{\codefont}[1]{\footnotesize{\texttt{#1}}\normalsize}

\newcommand{\totalproj}{57\xspace}

\newcommand{\totalvulproj}{40\xspace}
\newcommand{\totalsecproj}{17\xspace}

\newcommand{\projtnproj}{10\xspace}
\newcommand{\projtpproj}{38\xspace}

\newcommand{\projcorrectcases}{48\xspace}
\newcommand{\projpre}{84\%\xspace}
\newcommand{\projrecall}{95\%\xspace}
\newcommand{\projacc}{84\%\xspace}
\newcommand{\projf}{89\%\xspace}

\newcommand{\improvpre}{18\%\xspace}
\newcommand{\improvrec}{210\%\xspace}
\newcommand{\improvf}{123\%\xspace}
\newcommand{\improvacc}{100\%\xspace}

\newcommand{\rqacclo}{58\%\xspace}
\newcommand{\rqacchi}{84\%\xspace}
\newcommand{\rqflo}{64\%\xspace}
\newcommand{\rqfhi}{89\%\xspace}

\title{\tool: PoV-Guided Vulnerable API Usage Detection in Large Codebases via LLMs}
\titlerunning{\tool: PoV-Guided Vulnerable API Usage Detection}

\author{Quanzhi Fu$^{\star}$}{Virginia Tech, USA}{quanzhif@vt.edu}{}{}
\author{Wang Lingxiang$^{\star}$ }{Independent Researcher}{lingxiang.wang.2016@gmail.com}{}{}
\author{Wenjia Song}{Virginia Tech, USA}{wenjia7@vt.edu}{}{}
\author{Gelei Deng}{Nanyang Technological University, Singapore}{gdeng003@e.ntu.edu.sg}{}{}
\author{Yi Liu}{Nanyang Technological University, Singapore}{yi009@e.ntu.edu.sg}{}{}
\author{Dan Williams}{Virginia Tech, USA}{djwillia@vt.edu}{}{}
\author{Ying Zhang$^{\dagger}$}{Wake Forest University, USA}{ying.zhang@wfu.edu}{}{}
\begingroup
  \footnotetext{These authors contributed equally.}
  \footnotetext{Corresponding author.}
\endgroup
\authorrunning{Q. Fu, WL. Wang, W. Song, G. Deng, Y. Liu, D. Williams, and Y. Zhang}

\Copyright{Quanzhi Fu, Wang Lingxiang, Wenjia Song, Gelei Deng, Yi Liu, Dan Williams, and Ying Zhang}



\category{Technical Track Paper}       
\relatedversion{} 

\category{Technical Track Paper}
\EventEditors{Robert Feldt, Maria Paasivaara, Daniel Mendez, Stefan Wagner, and Marvin Mu\~{n}oz Bar\'{o}n}
\EventNoEds{5}
\EventLongTitle{20th International Symposium on Empirical Software Engineering and Measurement (ESEM 2026)}
\EventShortTitle{ESEM 2026}
\EventAcronym{ESEM}
\EventYear{2026}
\EventDate{October 8--9, 2026}
\EventLocation{Munich, Germany}
\EventLogo{}
\SeriesVolume{394}
\ArticleNo{1}

\begin{document}

\maketitle

\begin{abstract}


The integration of open-source libraries in Java development introduces severe security risks through vulnerable APIs. Existing program analysis 
and deep learning tools face the challenges of capturing inter-procedural vulnerability semantics at scale. While LLMs show promise for semantic reasoning, they cannot handle large codebases due to context limits and a lack of vulnerability-specific understanding to determine exploitability.
We present \tool, an LLM-powered framework for detecting vulnerable API usage through three core components. First, semantic-preserving AST-based fragmentation partitions large codebases while maintaining 
syntactic completeness within LLM context window limits. Second, vulnerability-aware multi-agent RAG traces relevant program context across these fragments, iteratively assembling security-critical context spanning functions and files. Third, PoV-guided semantic reasoning leverages Proof-of-Vulnerability tests that encode precise triggering conditions and exploitation mechanics for vulnerability determination. \tool achieves \projpre precision, \projrecall recall, \projacc accuracy, and \projf F1-score on \totalproj real-world Java applications, outperforming state-of-the-art tools. Our results show that vulnerability detection requires specialized architectural innovations beyond generic LLM applications.

\end{abstract}

\section{Introduction}\label{sec:intro}


Developers rely on many open-source libraries (Lib) and their application programming interface (API) to accelerate development. However, APIs that contain vulnerabilities can be exploited by hackers, turning widely used libraries into powerful attack vectors~\cite{wu2023understanding}. Attackers can introduce or discover exploitable vulnerabilities in Libs through (1) contributing flawed code to open-source projects, (2) inspecting library implementation, or (3) consulting publicly available vulnerability databases like Common Vulnerabilities and Exposures (CVE)~\cite{Zhang2025}. By crafting malicious inputs and propagating them from Apps to \textbf{vulnerable APIs}, attackers can compromise Applications (Apps) as long as the underlying Libs accept such inputs. The Open Web Application Security Project (OWASP) ranked ``software supply chain failures'' as the third most critical security risk in 2025~\cite{a03-owasp-2025}. 


To mitigate these security risks, existing software composition analysis (SCA) tools~\cite{ponta2020detection, zhang2024does, Li2021pdgraph, steenhoek2024dataflow, dependencychecker, snyk-dependency-check} identify Apps affected by vulnerable API usage via scanning code bases for (1) dependencies on vulnerable Lib versions and/or (2) calling vulnerable APIs in Libs. 
However, these tools suffer from low accuracy for two reasons. First, dependency-based analysis is overly coarse-grained and lacks context: an App may depend on a Lib that contains vulnerable APIs, yet never invokes those methods and therefore remains unaffected. 
As noted previously, ``\emph{SCA without context generates too many false positives, this results in your team wasting time chasing ghosts}~\cite{the-scariest}''. Second, SCA tools overestimate feasible attack paths at scale due to imprecise code semantic reasoning. VAScanner~\cite{zhang2024does}, for example, flags a vulnerability whenever a vulnerable API invocation is reachable along an approximate call chain, without verifying whether the usage is actually exploitable in the calling context.

Large Language Models (LLMs) have emerged as a promising alternative solution for vulnerability detection~\cite{zhou2024large, secllmholmes}. They offer semantic reasoning over code~\cite{ding2024semcoder} and can be guided through prompt engineering without large labeled datasets. 
Nevertheless, applying LLMs to detect vulnerable API usage in Apps has two challenges:  
\textbf{\textit{Challenge 1: Maintaining inter-procedural context.}} LLMs struggle to track repository-level context across deep function hierarchies~\cite{lu2024grace}. Since vulnerable API usage frequently spans multiple files connected by complex call chains, accurate detection requires preserving inter-procedural context. \textbf{\textit{Challenge 2: Reasoning about vulnerability-specific semantics.}} Trained on general-purpose corpora, LLMs lack the specialized security knowledge needed to reason about vulnerable API usage. Without vulnerability-oriented guidance, they misinterpret program semantics and produce unreliable detections.

In this paper, we introduce \tool, a high-accuracy framework for detecting vulnerable API usage in Apps using LLMs. Given a vulnerable API from an upstream Lib and an App's source code, \tool determines whether the API is exploitable from a public entry point in the app. \tool combines program analysis with LLM-based reasoning through two key designs. To maintain repo-level program context (\textbf{\textit{Challenge 1}}), we design a multi-agent pipeline in which specialized LLM agents iteratively retrieve missing security-critical context from AST-aligned code blocks and progressively assemble code fragments across functions/ files until the program context from vulnerable API invocation to public entry points is captured. To guide vulnerability-specific semantic reasoning (\textbf{\textit{Challenge 2}}), we leverage Proof-of-Vulnerability (PoV) tests from upstream Libs. PoV tests encode precise triggering conditions, exploitation mechanics, and observable impacts. \tool extracts these semantics from PoV tests and applies them to the repository-level context to determine whether the vulnerability is reachable and exploitable from public entry points.



We evaluate \tool on \totalproj real-world Java projects to assess its detection effectiveness. \tool achieves a precision of \projpre, a recall of \projrecall, an F1-score of \projf, and an accuracy of \projacc. Compared with five state-of-the-art (SOTA) tools, two based on program analysis (Eclipse Steady~\cite{ponta2020detection} and VAScanner~\cite{zhang2024does}), two DL-based approaches (DeepDFA~\cite{steenhoek2024dataflow} and LineVul~\cite{fu2022linevul}), and one LLM-agent based approach (Codex~\cite{openai_codex_cli}), \tool demonstrates substantial improvements, achieving at least \improvpre, \improvrec, \improvf, and \improvacc higher performance, respectively, across four evaluation metrics. To assess robustness, we further evaluate \tool varying configurations of embedding models, LLMs, and code fragmentation granularities. Across these settings,  \tool maintains better performance against SOTAs, with F1 scores ranging from \rqflo to \rqfhi and accuracy between \rqacclo and \rqacchi, demonstrating effectiveness across different models and code fragment sizes.

Our research contributions are summarized as follows:
\begin{itemize}
\item We propose \tool, an LLM-powered approach for vulnerable API usage detection in large-scale real-world applications, which overcomes LLMs’ context window limitations through an agent-based RAG that iteratively retrieves inter-procedural code fragments across multiple files.
\item We evaluated \tool on \totalproj real-world Java projects, demonstrating that it outperforms five SOTA tools across all metrics.

\item We conducted an extensive evaluation of different large language models (LLMs) and embedding models for vulnerability detection in code segments of varying lengths. We observed that for vulnerability detection, with OpenAI embeddings and \claud, consistently demonstrated higher performance compared to other tested models across all metrics.



\end{itemize}

Our code and dataset is available at: \url{https://github.com/ying-selab/CognixShield}.

\section{Background and Motivation Example}
\label{sec:background}

In this section, we first present a motivating example that illustrates the challenges of accurately detecting vulnerable API usages in large-scale software projects. We then provide the background on our system design.

\subsection{Motivation Example} \label{sec:example}


We use a concrete example to illustrate how the same vulnerable API can be exploitable in one application but not in another.
Consider a vulnerability analyst, Alex, who is assigned to audit two large software applications, Apache Kylin~\cite{apache_kylin_user_service} and Alibaba Nacos~\cite{nacos}, for potential risks caused by a CVE-2020-5408.
CVE-2020-5408 reveals a crypto failure in versions 4.2.x prior to 4.2.16. Specifically, the \codefont{BCryptPasswordEncoder.encode} method throws an \codefont{NullPointerException} when given a \codefont{null} input. 
This vulnerability was later patched in version 5.4.0-M1 by properly handling null inputs with an \codefont{IllegalArgumentException} as shown in Listing~\ref{lst:pov}.

\begin{minipage}{\linewidth}
\begin{MintedVerbatim}[commandchars=\\\{\},codes={\catcode`\$=3\catcode`\^=7\catcode`\_=8\relax},fontsize=\scriptsize,numbers=left,numberblanklines=true,breaklines=true,frame=lines,xleftmargin=1.0em,numbersep=3pt]
public String encode(CharSequence rawPassword) \PYGZob{}
\PYG{g+gi}{+ if (rawPassword == null) \PYGZob{}}
\PYG{g+gi}{+   throw new IllegalArgumentException(\PYGZdq{}rawPassword + cannot be null\PYGZdq{});\PYGZcb{}}
		...
\PYG{+w}{ } return BCrypt.hashpw(rawPassword.toString(), salt);\PYGZcb{}

//unit test for NPE
\PYG{g+gi}{+ @Test(expected = IllegalArgumentException.class)}
\PYG{g+gi}{+ public void encodeNullRawPassword() \PYGZob{}}
\PYG{g+gi}{+    BCryptPasswordEncoder encoder = new BCryptPasswordEncoder();}
\PYG{g+gi}{+    encoder.encode(null); \PYGZcb{}}
\end{MintedVerbatim}
\vspace{-1em}
\captionof{listing}{Simplified Fix Commit for CVE-2020-5408.~\cite{springsecurity_commit_d1909ec}}
\label{lst:pov}
\end{minipage}

Both Apache Kylin and Alibaba Nacos depend on the vulnerable library version and contain large codebases with over 5,000 source files and 500K lines of code. Alex needs to determine whether either application is actually exploitable due to this vulnerable API usage.
As shown in Listing~\ref{lst:vulusage}, Nacos invokes
\codefont{BCryptPasswordEncoder.encode} within \codefont{PasswordEncoderUtil.encode} without a null check (line 11), so an untrusted \codefont{null} can propagate through \codefont{createUser} (line 4) and trigger an application crash, enabling denial-of-service attacks. Conversely, as shown in Listing~\ref{lst:fp-case}, Apache Kylin~\cite{apache_kylin_user_service} validates all passwords via \codefont{checkPasswordLength} before encoding, ensuring null cannot reach the vulnerable API. Thus, despite using the same vulnerable library, Kylin is not exploitable.

\begin{figure}[hbtp]
\begin{minipage}[t]{0.47\linewidth}
\textbf{(a)} Nacos (vulnerable)
\begin{MintedVerbatim}[commandchars=\\\{\},fontsize=\scriptsize,numbers=left,breaklines=true,frame=lines,xleftmargin=1.0em,numbersep=3pt]
\PYG{k+kd}{public}\PYG{+w}{ }\PYG{n}{Object}\PYG{+w}{ }\PYG{n+nf}{createUser}\PYG{p}{(}\PYG{n}{String}\PYG{+w}{ }\PYG{n}{username}\PYG{p}{,}\PYG{+w}{ }\PYG{n}{String}\PYG{+w}{ }\PYG{n}{password}\PYG{p}{)}\PYG{+w}{ }\PYG{p}{\PYGZob{}}
\PYG{+w}{  }\PYG{p}{...}
\PYG{+w}{  }\PYG{n}{userDetailsService}\PYG{p}{.}\PYG{n+na}{createUser}\PYG{p}{(}\PYG{n}{username}\PYG{p}{,}
\PYG{+w}{  }\PYG{n}{PasswordEncoderUtil}\PYG{p}{.}\PYG{n+na}{encode}\PYG{p}{(}\PYG{n}{password}\PYG{p}{));}
\PYG{+w}{  }\PYG{k}{return}\PYG{+w}{ }\PYG{n}{RestResultUtils}\PYG{p}{.}\PYG{n+na}{success}\PYG{p}{(}\PYG{l+s}{\PYGZdq{}create user ok!\PYGZdq{}}\PYG{p}{);}
\PYG{p}{\PYGZcb{}}

\PYG{k+kd}{public}\PYG{+w}{ }\PYG{k+kd}{class} \PYG{n+nc}{PasswordEncoderUtil}\PYG{+w}{ }\PYG{p}{\PYGZob{}}
\PYG{+w}{  }\PYG{p}{...}
\PYG{+w}{  }\PYG{k+kd}{public}\PYG{+w}{ }\PYG{k+kd}{static}\PYG{+w}{ }\PYG{n}{String}\PYG{+w}{ }\PYG{n+nf}{encode}\PYG{p}{(}\PYG{n}{String}\PYG{+w}{ }\PYG{n}{raw}\PYG{p}{)}\PYG{+w}{ }\PYG{p}{\PYGZob{}}
\PYG{+w}{    }\PYG{k}{return}\PYG{+w}{ }\PYG{k}{new}\PYG{+w}{ }\PYG{n}{BCryptPasswordEncoder}\PYG{p}{().}\PYG{n+na}{encode}\PYG{p}{(}\PYG{n}{raw}\PYG{p}{);\PYGZcb{}}
\PYG{p}{\PYGZcb{}}
\end{MintedVerbatim}
\end{minipage}
\hfill
\begin{minipage}[t]{0.50\linewidth}
\textbf{(b)} Kylin (secure)
\begin{MintedVerbatim}[commandchars=\\\{\},fontsize=\scriptsize,numbers=left,breaklines=true,frame=lines,xleftmargin=1.0em,numbersep=3pt]
\PYG{k+kd}{public}\PYG{+w}{ }\PYG{n}{EnvelopeResponse}\PYG{+w}{ }\PYG{n+nf}{save}\PYG{p}{(}
\PYG{+w}{    }\PYG{n+nd}{@RequestBody}\PYG{+w}{ }\PYG{n}{PasswdChangeRequest}\PYG{+w}{ }\PYG{n}{user}\PYG{p}{)}\PYG{+w}{ }\PYG{p}{\PYGZob{}}
\PYG{+w}{  }\PYG{p}{...}
\PYG{+w}{  }\PYG{n}{checkNewPwdRule}\PYG{p}{(}\PYG{n}{user}\PYG{p}{.}\PYG{n+na}{getNewPassword}\PYG{p}{());}
\PYG{+w}{  }\PYG{k}{if}\PYG{+w}{ }\PYG{p}{(}\PYG{n}{existing}\PYG{+w}{ }\PYG{o}{!=}\PYG{+w}{ }\PYG{k+kc}{null}\PYG{p}{)}\PYG{+w}{ }\PYG{p}{\PYGZob{}}
\PYG{+w}{    }\PYG{n}{existing}\PYG{p}{.}\PYG{n+na}{setPassword}\PYG{p}{(}
\PYG{+w}{      }\PYG{n}{pwdEncode}\PYG{p}{(}\PYG{n}{user}\PYG{p}{.}\PYG{n+na}{getNewPassword}\PYG{p}{()));}\PYG{+w}{ }\PYG{p}{\PYGZcb{}\PYGZcb{}}

\PYG{k+kd}{private}\PYG{+w}{ }\PYG{k+kt}{void}\PYG{+w}{ }\PYG{n+nf}{checkNewPwdRule}\PYG{p}{(}\PYG{n}{String}\PYG{+w}{ }\PYG{n}{newPwd}\PYG{p}{)}\PYG{+w}{ }\PYG{p}{\PYGZob{}}
\PYG{+w}{  }\PYG{k}{if}\PYG{+w}{ }\PYG{p}{(}\PYG{o}{!}\PYG{n}{checkPasswordLength}\PYG{p}{(}\PYG{n}{newPwd}\PYG{p}{))}\PYG{+w}{ }\PYG{p}{\PYGZob{}...\PYGZcb{}}\PYG{+w}{ }\PYG{p}{\PYGZcb{}}

\PYG{k+kd}{private}\PYG{+w}{ }\PYG{k+kt}{boolean}\PYG{+w}{ }\PYG{n+nf}{checkPasswordLength}\PYG{p}{(}\PYG{n}{String}\PYG{+w}{ }\PYG{n}{password}\PYG{p}{)}\PYG{+w}{ }\PYG{p}{\PYGZob{}}
\PYG{+w}{  }\PYG{k}{return}\PYG{+w}{ }\PYG{o}{!}\PYG{p}{(}\PYG{n}{password}\PYG{+w}{ }\PYG{o}{==}\PYG{+w}{ }\PYG{k+kc}{null}
\PYG{+w}{        }\PYG{o}{||}\PYG{+w}{ }\PYG{n}{password}\PYG{p}{.}\PYG{n+na}{length}\PYG{p}{()}\PYG{+w}{ }\PYG{o}{\PYGZlt{}}\PYG{+w}{ }\PYG{l+m+mi}{8}\PYG{p}{);}\PYG{+w}{ }\PYG{p}{\PYGZcb{}}
\end{MintedVerbatim}
\end{minipage}
\captionof{listing}{Usage of \codefont{BCryptPasswordEncoder.encode}: (a)~vulnerable usage in Nacos~\cite{nacos} — \codefont{null} propagates unchecked to the API; (b)~secure usage in Kylin~\cite{apache_kylin_user_controller} — \codefont{checkPasswordLength} rejects \codefont{null} before encoding.}
\label{lst:vulusage}\label{lst:fp-case}
\vspace{-1em}
\end{figure}
Given the scale of these projects, manual code review is impractical, so Alex must rely on automated tools. Programming analysis tools~\cite{ponta2020detection, vascanner-github} encounter scalability challenges to accurately report vulnerable API usage. DL-based approaches~\cite{fu2022linevul, steenhoek2024dataflow} may not fully capture vulnerable API usage in Nacos, or overlook safeguard checks in Kylin.

While LLMs provide enhanced semantic reasoning capabilities, applying them directly to large-scale applications faces similar limitations. Their finite context window makes it impractical to analyze entire large codebases~\cite{ding2024semcoder}. In our preliminary experiments (detailed in Section~\ref{sec:rq3}), we found that the majority of real-world projects in our dataset exceed the context limits of the tested LLMs. Without an effective mechanism to selectively capture code relevant to vulnerable API usages and guidance for vulnerability-specific semantic reasoning, repo-wide analysis remains infeasible.

\subsection{Retrieval-Augmented Generation}
RAG~\cite{lewis2020retrieval} is a widely used framework that combines information retrieval with LLMs. A typical RAG architecture consists of a retriever and a generator: \textsc{retriever} searches an external knowledge source and the \textsc{generator} integrates
the retrieved information into its prompt to produce a more context-aware response. The general workflow involves: (1) embedding the knowledge base into a searchable index~\cite{openai_embeddings, qdrant}, (2) generating a query from the task prompt, (3) retrieving relevant entries, and (4) augmenting the prompt with these retrieved results before final response generation. An LLM agent is an autonomous system built on top of the LLM (e.g., GPT-4o~\cite{openai_gpt4o}, Claude~\cite{anthropic_claude_sonnet}). When combined with RAG, such agents form agentic RAG~\cite{singh2025agentic}, where the model dynamically refines its retrieval queries based on partial context (e.g.,
  reasoning about what information is missing) and performs multiple retrieval iterations. 

\section{Methodology}

\subsection{Overview}

\begin{figure*}[hbtp]
    \centering
    \includegraphics[width=0.80\textwidth]{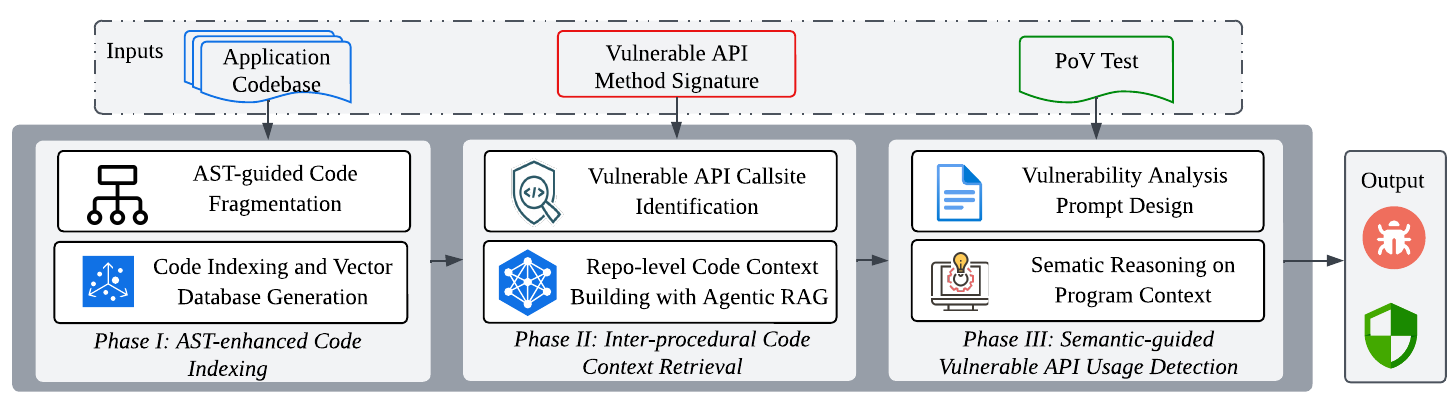}
    \caption{The Overview of \tool}
    \label{fig:overview}
    \vspace{-1em}
\end{figure*}

The overall workflow of \tool is depicted in Figure~\ref{fig:overview}. Given an application that depends on a vulnerable library version, the corresponding vulnerable API, and an available PoV patch, \tool analyzes whether the application context can make that API usage exploitable. The workflow is organized around the two challenges in the introduction: Phase~I maintains inter-procedural context for Challenge~1, and Phase~II applies vulnerability-specific semantic reasoning for Challenge~2.

\textit{Phase I: maintaining inter-procedural context.} \tool first parses the application source code with AST-guided structural segmentation, partitions the repository into semantically self-contained code blocks, and stores each block with its metadata and code embedding. This AST-enhanced code indexing makes repository-scale code searchable while preserving method, class, and file boundaries needed to recover caller/callee relationships around vulnerable library API invocations. Starting from each vulnerable API query, \tool then drives an Agentic RAG loop that iteratively retrieves and filters security-relevant program context, assembling sufficient repo-level inter-procedural context for the downstream exploitability decision (Section~\ref{sec:phase-i-context-retrieval}).

\textit{Phase II: reasoning about vulnerability-specific semantics.} Given the retrieved inter-procedural context and the PoV patch, \tool extracts PoV-derived vulnerability semantics, including trigger conditions, patch constraints, exploit mechanics, and observable impact. It then checks whether those conditions can propagate from public entry points through application data and control flows to the vulnerable API invocation. The output is an application-level decision that distinguishes exploitable vulnerable API usages from safe usages of the same library API.

\subsection{Phase I: Maintaining Repo-level Inter-procedural Context}

\subsubsection{AST-enhanced Code Indexing}
\label{sec:phase-i-context-maintenance}

Phase~I addresses Challenge~1 by turning the application repository into a searchable context index for vulnerable API usage analysis. A local invocation of a vulnerable library API rarely carries all the information needed to judge whether the usage is safe: the relevant caller, helper callee, field definition, object construction site, or guarding check often lives in another file, and the security-relevant data and control constraints span the structural units that the source code already exposes. The index must therefore preserve these program-defined units so that callers, callees, cross-file definitions, and data/control constraints remain reconnectable during retrieval. \tool builds the index from the program's AST rather than flattening the repository into arbitrary token chunks.
The root node (\texttt{compilationUnit}\footnote{We use the terms defined in Tree-sitter\cite{tree_sitter_java} for ease of understanding.}) corresponds to a source file, and each child node denotes a syntactic element such as a class, method, or statement. Each node contains metadata, including node type and source code location, and hierarchical relationships. Parent nodes structurally encompass their children; for example, a \texttt{classDeclaration} node encapsulates modifiers, fields, methods, constructors, and other elements defined at the class level within the class body.

To preserve semantically self-contained and contextually relevant code blocks, we leverage the AST to guide code partitioning.
We convert the application source code into an AST forest as
$G = \{g_1, g_2, \dots, g_n\}$, where each $g_i$ corresponds to a \texttt{compilationUnit}.
To obtain contextually relevant code blocks, we perform top-down, structure-guided fragmentation on each compilation unit. If a compilation unit is smaller than the predefined threshold $\theta$, it is preserved as a whole. Otherwise, oversized nodes are recursively divided into child-nodes.
Recursion stops when a node falls below $\theta$ or reaches a leaf with no further structural decomposition. For instance, a \texttt{compilationUnit} can be fragmented as \texttt{ImportDeclaration} and \texttt{classDeclaration}.
This AST-guided structural segmentation keeps repository code aligned with source-level method, class, and file boundaries, so later retrieval can reconnect vulnerable API invocations with surrounding caller/callee paths and security-relevant data/control constraints.


Following the AST-enhanced code segmentation, \tool converts
each code block $b_i$ in $B = \{b_1, b_2, \dots, b_m\}$
into a fixed-length dense vector representation using pre-trained encoders. These embeddings capture the semantics and functionality of the fragments, positioning semantically similar blocks closer together in the vector space, even if their syntactic forms differ. For instance, the following two snippets perform the same task of summing elements despite having different token sequences: \\
\indent \codefont{a) for (int i = 0; i < n; i++) { sum += arr[i]; }}\\
\indent \codefont{b) total = Arrays.stream(list).sum(); }\\
Although these snippets use different tokens and syntax, once embedded, they yield a high similarity score because they perform the same functionality.   
Within Phase~I, these embeddings serve as retrieval keys rather than standalone predictions.
\tool therefore stores all processed data in a database $D=\{d_1,d_2,\ldots,d_m\}$, where each entry $d_i=(b_i, v_i)$ consists of the original code block $b_i$, which includes the source snippet and associated metadata 
, and its corresponding vector embedding $v_i$.
This database becomes the searchable substrate for the subsequent Phase~I context-retrieval step, where semantically related fragments are assembled into sufficient repo-level inter-procedural context before Phase~II performs vulnerability-specific semantic reasoning.

\subsubsection{Agentic RAG for Inter-procedural Context Retrieval}
\label{sec:phase-i-context-retrieval}



To address Challenge~1 for context retrieval, \tool assembles the required inter-procedural context through an Agentic RAG framework built on LLM-based reflection~\cite{asai2023selfrag}, consisting of two collaborating LLM-based agents: a \textit{Reflector} and a \textit{Grader}. At each iteration, the \textit{Reflector} reasons about missing caller/callee, data-construction, validation, or control-flow links and formulates focused embedding queries against the AST-enhanced vector database $D$ described above, while the \textit{Grader} retains only fragments with semantic relevance to the vulnerable API usage.

The loop iterates over a set of candidate vulnerable API invocation sites.
Given a vulnerable library API ($API_{vul}$), \tool first locates its invocation sites in the application by matching the method signature $API_{vul}$, forming the initial statement-level candidates $C_{candidate}=\{c_1,c_2,\ldots,c_n\}$.

\begin{figure}[t]
    \centering
    \includegraphics[width=0.7\textwidth]{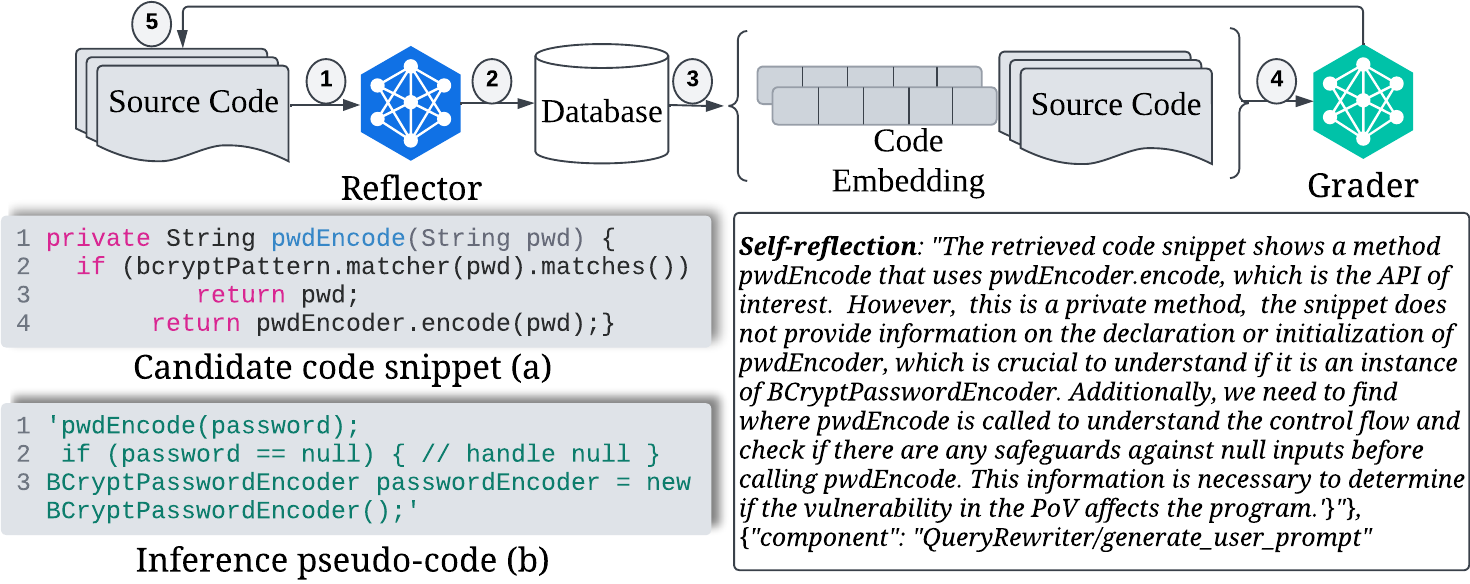}
    \caption{Interaction process between \textit{Reflector} and \textit{Grader} agents on an exemplar candidate for the \codefont{BCryptPasswordEncoder.encode()} in Apache Kylin.}
    \label{fig:candidate-context-retrive}
    \vspace{-1em}
\end{figure}

Given these candidates, the \textit{Reflector} and \textit{Grader} interact as follows:
\begin{itemize}
    \item \textit{Self-Reflection:} For each candidate $c_i \in C_{candidate}$, the \textit{Reflector} evaluates whether the current fragment set provides sufficient inter-procedural context for analyzing the vulnerable API usage.
    It issues the reflection query
    $Q(c_i, API_{vul}) \rightarrow (\{\texttt{Complete|Incomplete}\}, \text{reason})$.
    A \texttt{Complete} response indicates that the current fragments contain the relevant elements needed for the Phase~II Expert agent to reason about the invocation path, while an \texttt{Incomplete} response explains which related code is missing and guides subsequent inference.
    The missing link may be a caller, callee, public entry point, object construction statement, validation check, or control-flow constraint.
    For example, the \textit{Reflector} examines candidate snippet~(a) in Figure~\ref{fig:candidate-context-retrive}, where the vulnerable API \codefont{pwdEncoder.encode()} is invoked inside a private method.
    This fragment alone lacks sufficient context because it does not show how the password value is constructed or guarded before the call;
    the \textit{Reflector} therefore returns an \texttt{Incomplete} verdict naming the missing caller, data-construction, and validation links.
    \item \textit{Code-inference \& Retrieval:} When the context is \texttt{Incomplete}, the \textit{Reflector}
    formulates a focused retrieval query for the missing element identified by the reflection verdict.
    The query takes the form of an inferred code sketch that approximates the missing element's expected signature and surrounding scope.
    As shown in pseudo-code~(b), the \textit{Reflector} sketches how \codefont{pwdEncode()} may be invoked by its callers and whether the \codefont{password} parameter is checked for null before reaching \codefont{pwdEncoder.encode()}.
    This inferred information drives an embedding-based search over the repository vector database $D$, allowing \tool to retrieve AST-aligned fragments from across the repository.

    \item \textit{Grading}: The \textit{Grader} evaluates the raw retrieval results at the source level and verifies their semantic relevance to the vulnerable API usage.
    A retained snippet must extend the security-relevant inter-procedural chain, for example by establishing a caller/callee relationship, defining data passed into the API, implementing a validation check, or constraining the control path to the invocation.
    Snippets that share lexical similarity but do not affect the vulnerable API usage are filtered out.
    Validated snippets are merged into the candidate set: $C_{candidate_i} = C_{candidate_i} \cup C_{validated}$.
    This step mitigates false positives that embedding similarity alone may introduce.
\end{itemize}
The process repeats until the \textit{Reflector} judges that the current fragment set provides sufficient inter-procedural context for vulnerability reasoning, or until retrieval produces no additional non-duplicated validated fragments.
The process is shown at Algorithm~\ref{alg:phase-i-context-retrieval}.

\begin{algorithm}[htbp]
\scriptsize
\caption{Iterative Phase-I Inter-procedural Context Retrieval}
\label{alg:phase-i-context-retrieval}
\begin{algorithmic}[1]
    \Require{$C_{candidates} = \{c_1, c_2, \ldots, c_n\}$, $API_{vul}$}
    \Statex \hspace*{-2em} \textbf{Auxiliary Functions:}
    \Statex \hspace*{-1em} \texttt{Search(code, scope)} $\to$ {$C$}: Search for similar code based on embeddings
    \Statex \hspace*{-1em} \texttt{Reflector\_evaluate($C$, $API_{vul}$)} $\to$ (hasSufficientContext, missingContext): Determines whether $C$ provides sufficient inter-procedural context for analyzing $API_{vul}$
    \Statex \hspace*{-1em} \texttt{Grader\_filter($C_{raw}$, context)} $\to$ $C_{filtered}$: Filters out irrelevant snippets from raw search results based on context relevance
    \Statex \hspace*{-1em} \texttt{Reflector\_Inf(missingContext)} $\to$ $(c_{missing}, scope)$: Generates missing code snippet and structural constraints

    \For{each $c_i \in C_{candidates}$}
        \State $C' \leftarrow \{c_i\}$
        \State $(hasSufficientContext, missingContext) \leftarrow \texttt{Reflector\_evaluate}(C', API_{vul})$
        \While{not $hasSufficientContext$}
            \State $(c_{missing}, scope) \leftarrow \texttt{Reflector\_Inf}(missingContext)$
            \State $C'_{raw} \leftarrow \texttt{Search}(c_{missing}, scope)$
            \If{$C'_{raw} == \emptyset$}
                \State \textbf{break}
            \EndIf
            \State $C'_{candidate} \leftarrow \texttt{Grader\_filter}(C'_{raw}, C' \cup \{API_{vul}\})$
            \If{$C'_{candidate} == \emptyset$}
                \State \textbf{break}
            \EndIf
            \State $C' \leftarrow C' \cup C'_{candidate}$
            \State $C_{candidates} \leftarrow C_{candidates} \cup C'_{candidate}$
            \State $(hasSufficientContext, missingContext) \leftarrow \texttt{Reflector\_evaluate}(C', API_{vul})$
        \EndWhile
    \EndFor
\end{algorithmic}
\end{algorithm}

\subsection{PoV-guided Vulnerable API Usage Detection}
\label{sec:Phase II-2}

To address Challenge~2, \tool performs vulnerability-specific semantic reasoning over the inter-procedural context assembled in Phase~I.
After retrieving the sufficient inter-procedural context, we introduce an \textit{Expert} agent that applies PoV-derived vulnerability semantics to that context.
The \textit{Expert} receives a structured system prompt and standardized inputs/outputs to ensure reliable reasoning.

The system prompt 
is specifically designed to enable context-aware detection of vulnerable API usage. Unlike generic prompts that merely classify code as vulnerable or secure, our prompt instructs a two-stage reasoning process:
(i) \textbf{PoV-derived vulnerability semantics extraction}, where the model extracts trigger conditions, patch constraints, exploit mechanics, and observable impact from the provided PoV;
and
(ii) \textbf{semantic propagation analysis}, where multiple related code fragments are analyzed collectively to deduce cross-method path or data constraints, determining whether the exploit condition can propagate to the vulnerable API invocation.
For each candidate $c_i$ in the set $C_i = \{c_1, c_2, \ldots, c_n\}$ from Phase~I, we construct a \textit{task-specific} guidance prompt to the \textit{Expert}:
\begin{mybox}{\textbf{\textit{Vulnerable API Usage Analysis Prompt}}}
\textit{You are an expert in Java vulnerability analysis. Your task is to determine if a project is affected by a library vulnerability. Given a project, a vulnerable API, and a library PoV, analyze whether the complete snippet set is affected by the vulnerable API Usage.}
\\
\texttt{Vulnerable API:}\\
\texttt{Library POV:}\\
\texttt{Related code snippet:} ...
\end{mybox}

Given this structured input, the \textit{Expert} executes the two stages above and produces an interpretable vulnerability detection result. For instance,
based on the code example in the Kylin project (Section~\ref{sec:example}), \tool analyzes whether variables \codefont{pwd} and calls to \codefont{pwdEncoder.encode(pwd)} can result in a null value reaching the vulnerable API \codefont{BCryptPassword\\Encoder.encode()}, as described in the PoV. The \textit{Expert} generates conclusions such as: \textit{"The code has multiple safeguards
preventing null passwords from reaching the
encoder."}, and the project is not vulnerable to the API usage.

Building on this analysis, \tool evaluates each candidate $c_i$ independently using its corresponding prompt $Q_i$, and the LLM returns a binary decision: $1$ (vulnerable) or $0$ (secure).
The application is classified as vulnerable if any candidate query is identified as exploitable; it is classified as secure only when all candidates are assessed as non-exploitable.

\section{Implementation}
\label{sec:implementation}
In phase I, we employ the tree-sitter~\cite{tree_sitter_java} to generate an AST representation of the codebase. Subsequently, our tool partition the codebase into code blocks via recursive AST decomposition with a $\theta = 2500$ token threshold, which are then serialized into a single JSON file. Each JSON object encapsulates essential metadata, including the filename, start line, end line, code snippet, and relevant contextual information. After fragmentation, each code block is embedded using the OpenAI text-embedding-3-small model~\cite{openai_embeddings}, which is widely used in industry AI coding tools, while its metadata fields are preserved. We store the resulting embeddings and metadata in Qdrant~\cite{qdrant}, a vector database, using cosine distance and a similarity threshold of 0.35 for retrieval. Both the \textit{Reflector} and \textit{Grader} use GPT-4o with temperature 0 to minimize output randomness. We use Tai-e~\cite{tai-e} to locate candidate invocation sites of $API_{vul}$ with the vulnerable API signature given as input, e.g., \codefont{com.alibaba.fastjson.JSON.parseObject} in the project's compiled bytecode to initialize the context retrieval; when the signature names an overloaded API, the sites of all overloads are included as candidates.  In Phase~II, the \textit{Expert} agent is implemented as a GPT-4o-based LLM agent with temperature 0. 

\section{Evaluation}


Our evaluation aims to answer the following Research Questions (RQs):

\begin{itemize}
\item \textbf{RQ1 (Effectiveness):} \emph{How effectively does \tool identify vulnerable API usage in large projects?}
\item \textbf{RQ2 (Component Contribution):} \emph{How does each of \tool's three core components contribute to its effectiveness?}
\item \textbf{RQ\oldx{3} (Baseline Comparison):} \emph{How well does \tool compare with SOTA solutions?}
\item \textbf{RQ\oldx{4}(Robustness):} \emph{How robust is \tool's vulnerability detection capability across different LLM-related configurations?}
\end{itemize}


\noindent\textbf{Datasets.}
\label{sec:dataset}
To evaluate our tool, we construct the evaluation dataset based on the following sources.
The foundation of our dataset comes from Transfer~\cite{kang2022test}, a third-party benchmark that provides paired Java projects and their vulnerable libraries. Due to compatibility issues in Transfer, many vulnerable libraries lack corresponding client projects. To enlarge the dataset, we search on GitHub using vulnerable library names as keywords and retrieve the top 100 Java projects for each vulnerable library. We then filter these projects based on vulnerable library version and actual API invocation. The resulting dataset includes 57 projects corresponding to 26 distinct vulnerabilities.
To establish ground truth, two security experts independently analyze and cross-verify the exploitability of all projects in the dataset. Among them, 40 applications are confirmed to be vulnerable where PoV tests are successfully generated to demonstrate vulnerabilities caused by the vulnerable API usage in these applications. The remaining 17 projects are verified as unaffected.
The complete project list and labels are provided in the replication package (see Data Availability). On average, each project contains 1,685 Java files and 193,541 lines of code.

\begin{table}
  \caption{The precision, recall, F1-score and accuracy of tools measured based on \totalproj projects}\label{tab:tool-comparison}
 
  \setlength{\tabcolsep}{3pt}
  \footnotesize
  \begin{adjustbox}{max width=\linewidth}
  \begin{tabular}{l| rr|rrrr|rrrr|rrrr|rrrr|rrrr|rrrr|rrrr}
  \toprule
  \multirow{2}{*}{\textbf{CWE Category}} & \multicolumn{2}{c|}{\textbf{GT}} &\multicolumn{4}{c|}{\textbf{CognixShield(\%)}} & \multicolumn{4}{c|}{\textbf{NaiveLLM(\%)}} &
  \multicolumn{4}{c|}{\textbf{ Codex(\%)}} &\multicolumn{4}{c|}{\textbf{VAScanner(\%)}} &\multicolumn{4}{c|}{\textbf{Steady (\%)}} &\multicolumn{4}{c|}{\textbf{DeepDFA(\%)}}
  &\multicolumn{4}{c}{\textbf{LineVul(\%)}} \\ \cline{2-31}
  & \textbf{TP} &\textbf{TN} &
  \textbf{P} &\textbf{R} &\textbf{F}& \textbf{A} &
  \textbf{P} &\textbf{R} &\textbf{F}& \textbf{A} &
  \textbf{P} &\textbf{R} &\textbf{F}& \textbf{A} &
  \textbf{P} &\textbf{R} &\textbf{F}& \textbf{A} &
  \textbf{P} &\textbf{R} &\textbf{F} & \textbf{A} &
  \textbf{P} &\textbf{R} &\textbf{F}& \textbf{A} &
  \textbf{P} &\textbf{R} &\textbf{F}& \textbf{A} \\ \hline
  CWE-20 (13) &9&4 &89&89&89&85 &75&25&38&23 &82&100&90&85&75&100&86&75 &83&45&59&46 &50&22&31&31 &67&22&33&38 \\ \hline
  CWE-22 (6) &5&1 &83&100&91&83 &100&17&29&17 &80&80&80&67&0&-&-&0 &67&40&50&33 &100&20&33&33 &100&20&33&33 \\ \hline
  CWE-78 (7) &6&1 &100&100&100&100 &-&0&-&14 &83&83&83&71&100&100&100&100 &100&29&44&29 &50&17&25&14 &75&50&60&43 \\ \hline
  CWE-91 (3) &2&1 &67&100&80&67 &100&33&50&33 &67&100&80&67&100&100&100&100 &-&0&-&33 &-&0&-&33 &-&0&-&33 \\ \hline
  CWE-180 (3) &2&1 &67&100&80&67 &100&100&100&100 &67&100&80&67&100&33&50&33 &-&0&-&0 &100&50&67&67 &-&0&-&33 \\ \hline
  CWE-330 (4) &2&2 &100&100&100&100 &100&100&100&100 &-&0&-&50&0&-&-&0 &-&0&-&0 &50&50&50&50 &50&50&50&50 \\ \hline
  CWE-502 (4) &3&1 &100&100&100&100 &0&-&-&0 &100&33&50&50&-&-&-&- &67&100&80&67 &-&0&-&25 &0&0&-&0 \\ \hline
  CWE-611 (2) &2&0 &100&100&100&100 &100&100&100&100 &100&100&100&100&-&-&-&- &-&-&-&- &-&0&-&0 &100&50&67&50 \\ \hline
  CWE-770 (6) &4&2 &80&100&89&83 &100&100&100&100 &60&75&67&50&50&100&67&50 &50&20&29&17 &-&0&-&33 &100&75&86&83 \\ \hline
  CWE-787 (3) &1&2 &33&100&50&33 &-&-&-&- &33&100&50&33&-&-&-&100 &0&0&-&0 &-&0&-&67 &-&0&-&67 \\ \hline
  CWE-835 (4) &3&1 &100&100&100&100 &100&100&100&100 &75&100&86&75&50&33&40&25 &0&0&-&0 &-&0&-&25 &-&0&-&25 \\ \hline
  CWE-1025 (2) &1&1 &-&0&-&50 &-&-&-&- &-&0&-&0&-&-&-&100 &0&-&-&0 &-&0&-&50 &50&100&67&50 \\
  \hline
  \textbf{Overall (\totalproj)} & \textbf{40} &\textbf{17}
  & \textbf{84} &\textbf{95} & \textbf{89} & \textbf{84}
  &\textbf{84} &\textbf{28} &\textbf{42}& \textbf{30}
  &\textbf{73} &\textbf{80} &\textbf{76}& \textbf{65}
  &\textbf{64} &\textbf{23} &\textbf{33}& \textbf{25}
  &\textbf{60}&\textbf{30}&\textbf{40}& \textbf{28}
  &\textbf{60}&\textbf{15}&\textbf{33} & \textbf{24}
  &\textbf{71}&\textbf{30}&\textbf{42} & \textbf{42} \\ \bottomrule
  \multicolumn{27}{l}{``-'' means that a tool does not report vulnerability of API usage for applications, so the related precision and F-score values cannot be calculated. } \\
  \end{tabular}
  \end{adjustbox}
  \vspace{-2em}
  \end{table}

\noindent\textbf{Metrics.} Following prior work~\cite{Rahaman2019}, we use four metrics to evaluate \tool: 

\textbf{Precision (P):}  The ratio of correctly identified vulnerable projects to the total number of projects flagged as vulnerable by the tool.
$P = \frac{\text{\# of correctly identified vulnerable projects}}{\text{Total \# of projects flagged as vulnerable}}$
\normalsize

\textbf{Recall (R):} The ratio of correctly identified vulnerable projects to the total number of actual vulnerable projects in the ground truth dataset.
$R = \frac{\text{\# of correctly identified vulnerable projects}}{\text{Total \# of vulnerable projects in ground truth}}$

\emph{\textbf{F1-score (F1)}} is the harmonic mean of precision and recall; it reflects the trade-off between precision and recall. 
$F1 = \frac{2 \times P \times R}{P + R} $

\textbf{Accuracy (A):} The ratio of correctly classified projects to the total number of projects in the ground truth dataset.
$A = \frac{\text{\# of correctly classified projects}}{\text{Total \# of projects in ground truth}}$




\subsection{RQ1 (Effectiveness of \tool )}\label{sec:rq1}
We evaluate \tool on the \totalproj open-source Java projects using the same hyperparameters as Section~\ref{sec:implementation}. \tool analyzed \totalproj applications that invoke the vulnerable API and correctly classified \projcorrectcases of them~\ying{as shown in Table~\ref{tab:tool-comparison}}. Specifically, among the \totalvulproj projects confirmed to be vulnerable, \tool successfully identified vulnerabilities in \projtpproj of them. For the \totalsecproj projects verified as secure, \projtnproj were correctly classified as unaffected. Overall,  \tool achieves a P of \projpre, R of \projrecall, F1 of \projf, and A of \projacc. 
We then manually inspected the running logs on the evaluated applications to understand the effectiveness of \tool. 
The nine incorrect detections fall into three categories: 

1) \textit{\underline{Incomplete context retrieval.}} Oversized AST leaf nodes exceeded the fragment size limit, forcing semantic fragmentation that disrupted contextual reconstruction. Three false positives arose when large leaf nodes, like string literals, exceeded the token limit and could not fit within a single fragment. Since leaf nodes are syntactically atomic, \tool's hierarchical decomposition had to split them at arbitrary points, disrupting their semantic meaning. For instance, in Findbugs~\cite{findbugs_bugtreemodel}, a long statement in \codefont{branchOperations} was split mid-expression, preventing \tool from reconstructing the inter-procedural context. Such fragmentation makes it challenging for \tool to reconstruct the relevant context needed for accurate analysis.

2) \textit{\underline{Relevant code fragments missed.}}
\lx{Large, security-critical code blocks were often missed because their embeddings were dissimilar to the retrieval query.}
Four failures (2 FPs, 2 FNs) resulted from embedding-based retrieval, missing large but relevant code blocks.
For example, in the Badlion project~\cite{leicht2024badlion}, the 40-50 LOC \codefont{URIUtils} class containing the vulnerable API fell below the similarity threshold. Instead, \tool retrieved the API's library implementation, misinterpreting library code as application logic and producing a false positive.

3) \textit{\underline{Reasoning failures for vulnerability semantics.}}
Two false positives occurred when LLM reasoning introduced assumptions beyond those specified in the PoV. For example, in \codefont{ia-recruiter} application~\cite{BasicNameFinder}, the code hardcodes the ZIP file path to prevent directory traversal. However, \tool assumed attackers could replace the hardcoded file itself, incorrectly flagging it as vulnerable to CVE-2019-10094\cite{CVE-2019-10094} despite the PoV indicating otherwise.
While prompt engineering with additional constraints helps, LLMs inevitably generate assumptions beyond provided information, making these reasoning errors difficult to eliminate entirely.

\begin{adjustwidth} {0mm}{0mm}
\begin{tcolorbox}
[colback=gray!8, size=title, boxsep=1mm, colframe=black, before={\vskip1mm}, after={\vskip0mm}]
\textbf{Finding 1:}
\textit{\tool correctly classifies \projtpproj vulnerable and \projtnproj secure applications, achieving a precision of \projpre, recall of \projrecall, F1-score of \projf, and accuracy of \projacc across \totalproj applications. The unbalanced code structures can disrupt AST-guided fragmentation and degrade the quality of retrieved context.}
\end{tcolorbox}
\end{adjustwidth}

Upon further inspection, we observed that the embedding cost increases with project size because the entire codebase must be indexed. For example, the largest project, Hadoop (2.63M LoC), requires 19.8M embedding tokens and takes approximately 76 minutes to generate embeddings. However, the overall analysis cost is primarily determined by the number of vulnerable API invocations rather than project size. Among the two projects exceeding 1M LoC, Hadoop and Druid (1.21M LoC), our approach successfully analyzed both: Hadoop requires 339K input tokens and 7K output tokens for LLM analysis (approximately \$1.30), whereas Druid requires 2.4M input tokens and 360K output tokens (approximately \$10.02). This difference is mainly because Druid contains 28 vulnerable API invocations, compared to only 8 in Hadoop. On average, our approach used 1,772,500 tokens for embedding and 174,196 LLM tokens (88.2\% input and 11.8\% output), resulting in an average cost of approximately \$0.60 per project analysis.


\begin{adjustwidth} {0mm}{0mm}
\begin{tcolorbox}
[colback=gray!8, size=title, boxsep=1mm, colframe=black, before={\vskip1mm}, after={\vskip0mm}]
\textbf{Finding 2:}
\textit{\lx{\tool tool analyzes large-scale codebases at low cost, for an average of \$0.60 per project analysis. This cost is primarily determined by the number of vulnerable API invocations rather than project size, keeping \tool practical for multi-million-LoC repositories.}}
\end{tcolorbox}
\end{adjustwidth}

\subsection{RQ2 (Component Contribution)}\label{sec:rq2}

\textbf{Experiment setting.} To better understand the contribution of each component, we conducted an ablation study to evaluate the effectiveness of \tool under the following settings:
(1) w/o AST, where the AST-guided code segmentation is replaced with naive line-based chunking; and
(2) w/o RAG, where only the target function that invokes the vulnerable API is provided as context; and \newx{(3) w/o PoV, where the PoV test is not included.}

\begin{table}[htbp]
  \centering
  \caption{Ablation experiment results of \tool components }
  \label{tab:ablation}
  \scriptsize
  \begin{tabular}{lcccc}
    \toprule
    \textbf{Model} & \textbf{P (\%)} & \textbf{R (\%)} & \textbf{F(\%)} & \textbf{A(\%)} \\
    \midrule
    CognixShield & 84 & 95 & 89& 84  \\
    CognixShield w/o AST & 79 & 54 & 64 & 57 \\
    CognixShield w/o RAG & 70 & 53 & 60 & 50 \\
    \newx{CognixShield w/o PoV} & \newx{72} & \newx{83} & \newx{77} & \newx{65} \\

    \bottomrule
  \end{tabular}
\end{table}


As shown in Table~\ref{tab:ablation}, all three settings led to substantial performance degradation.
Specifically, without AST, recall dropped by 41\% and accuracy dropped by 27\%, while no RAG reduced recall by 42\% and accuracy by 34\%.  Excluding PoV guidance decreased both precision and recall by 12\%.
\ying{we observed that both AST-based indexing and RAG are critical for retrieving security-relevant inter-procedural context. 
Without AST-based indexing. \tool loses structural information needed to identify semantically coherent code fragments. For example, when analyzing CVE-2022-25845 in the \texttt{commerce} project, the vulnerability is exploitable only when \codefont{JSON.parseObject} deserializes malformed JSON strings. Without AST-based indexing, the retrieved code snippets consist primarily of isolated variable assignments and omit the surrounding control-flow structure.}

\textit {\textbf{Inter-procedural Context Retrieval.}}
\ying{We observed that \tool retrieves security-relevant context on demand, adapting its retrieval scope to the vulnerability under analysis rather than reconstructing the entire program. Across 47 projects, \tool retrieved cross-function and cross-file code fragments to establish exploitability, whereas for the remaining 10 projects, reasoning within a single function was sufficient.}
Listing~\ref{lst:vulnprompt-nacos} shows a simplified prompt automatically generated by \textit{Reflector} for the Nacos project.  \tool performed control-flow related code fragments retrieval across multiple files; it retrieved code blocks from the public interface \codefont{createUser} down to the vulnerable API invocation \codefont{PasswordEncoderUtil.encode}.
While in the Kylin project (Listing~\ref{lst:vulnprompt-kylin}), \tool gathered data constraints context instead of the entire control flow. It includes the \codefont{checkNewPwdRule} function, which has security-related logic for password validation. For both cases, the retrieved context enabled \tool to correctly identify the trigger condition, impact, and exploitation flow. When RAG is disabled, \tool analyzes only the local function containing the vulnerable API, preventing it from determining whether the invocation is reachable under exploitable conditions.
\ying{These findings indicate that accurate detection depends on retrieving security-relevant inter-procedural context rather than analyzing the function containing the vulnerable API invocation in isolation.}

\begin{figure}[hbtp]
\vspace{-1em}
\begin{minipage}[t]{0.47\linewidth}
\textbf{(a)} Nacos (vulnerable)
\begin{pygcolorbg}{lightgray!8}
\begin{MintedVerbatim}[commandchars=\\\{\},fontsize=\tiny,breaklines=true,breakanywhere=true,frame=single,framesep=2mm]
Related code snippet:
\PYGZhy{}\PYGZhy{}\PYGZhy{}\PYGZhy{}\PYGZhy{}\PYGZhy{}\PYGZhy{}\PYGZhy{}\PYGZhy{}\PYGZhy{}\PYGZhy{}\PYGZhy{}\PYGZhy{}\PYGZhy{}\PYGZhy{}\PYGZhy{}\PYGZhy{}\PYGZhy{}\PYGZhy{}\PYGZhy{}\PYGZhy{}
file\PYGZus{}path: PasswordEncoderUtil.java
Code:
public class PasswordEncoderUtil \PYGZob{}
  ...
  public static String encode(String raw) \PYGZob{}
    return new BCryptPasswordEncoder()
              .encode(raw);\PYGZcb{}\PYGZcb{}
\PYGZhy{}\PYGZhy{}\PYGZhy{}\PYGZhy{}\PYGZhy{}\PYGZhy{}\PYGZhy{}\PYGZhy{}\PYGZhy{}\PYGZhy{}\PYGZhy{}\PYGZhy{}\PYGZhy{}\PYGZhy{}\PYGZhy{}\PYGZhy{}\PYGZhy{}\PYGZhy{}\PYGZhy{}\PYGZhy{}\PYGZhy{}
file\PYGZus{}path: UserController.java
Code:
public Object createUser(
    String username, String password) \PYGZob{}
  ...
  userDetailsService.createUser(username,
    PasswordEncoderUtil.encode(password));
  return RestResultUtils.success(\PYGZdq{}ok!\PYGZdq{});\PYGZcb{}
\end{MintedVerbatim}
\end{pygcolorbg}
\end{minipage}
\hfill
\begin{minipage}[t]{0.50\linewidth}
\textbf{(b)} Kylin (secure)
\begin{pygcolorbg}{lightgray!8}
\begin{MintedVerbatim}[commandchars=\\\{\},fontsize=\tiny,breaklines=true,breakanywhere=true,frame=single,framesep=2mm]
Related code snippet:
\PYGZhy{}\PYGZhy{}\PYGZhy{}\PYGZhy{}\PYGZhy{}\PYGZhy{}\PYGZhy{}\PYGZhy{}\PYGZhy{}\PYGZhy{}\PYGZhy{}\PYGZhy{}\PYGZhy{}\PYGZhy{}\PYGZhy{}\PYGZhy{}\PYGZhy{}\PYGZhy{}\PYGZhy{}\PYGZhy{}\PYGZhy{}
file\PYGZus{}path: UserController.java, L252\PYGZhy{}260
Code:
private void checkNewPwdRule(String newPwd) \PYGZob{}
  if (!checkPasswordLength(newPwd)) \PYGZob{}...\PYGZcb{}\PYGZcb{}
\PYGZhy{}\PYGZhy{}\PYGZhy{}\PYGZhy{}\PYGZhy{}\PYGZhy{}\PYGZhy{}\PYGZhy{}\PYGZhy{}\PYGZhy{}\PYGZhy{}\PYGZhy{}\PYGZhy{}\PYGZhy{}\PYGZhy{}\PYGZhy{}\PYGZhy{}\PYGZhy{}\PYGZhy{}\PYGZhy{}\PYGZhy{}
file\PYGZus{}path: UserController.java, L262\PYGZhy{}264
Code:
private boolean checkPasswordLength(
    String password) \PYGZob{}
  return !(password == null
        || password.length() \PYGZlt{} 8);\PYGZcb{}
\PYGZhy{}\PYGZhy{}\PYGZhy{}\PYGZhy{}\PYGZhy{}\PYGZhy{}\PYGZhy{}\PYGZhy{}\PYGZhy{}\PYGZhy{}\PYGZhy{}\PYGZhy{}\PYGZhy{}\PYGZhy{}\PYGZhy{}\PYGZhy{}\PYGZhy{}\PYGZhy{}\PYGZhy{}\PYGZhy{}\PYGZhy{}
file\PYGZus{}path: UserController.java, L201\PYGZhy{}238
Code:
public EnvelopeResponse save(
    @RequestBody PasswdChangeRequest user) \PYGZob{}
  ...
  checkNewPwdRule(user.getNewPassword());
  if (existing != null) \PYGZob{}
    existing.setPassword(
      pwdEncode(user.getNewPassword()));\PYGZcb{}...\PYGZcb{}
\end{MintedVerbatim}
\end{pygcolorbg}
\end{minipage}
\captionof{listing}{Context retrieved by \textit{Reflector}: (a)~Nacos~\cite{nacos} — cross-file control-flow from \codefont{createUser} to the vulnerable API; (b)~Kylin~\cite{apache_kylin_user_controller} — data-constraint chain showing null validation before encoding.}
\label{lst:vulnprompt-nacos}\label{lst:vulnprompt-kylin}
\end{figure}

\textbf{\textit{PoV-guided Vulnerability Semantic Reasoning.}} 
With PoV-guided, 
\tool interprets the retrieved code context with vulnerability semantics (e.g.,  vulnerability’s triggering conditions) with the PoV guidance.  
For example, to analyze the Kylin and Nacos projects, \tool first deduces from the PoV test (Listing~\ref{lst:pov}, Section~\ref{sec:example}) that exploitation is only possible when a null password is passed to the \codefont{BCryptPasswordEncoder.encoder()} method. Leveraging this semantic understanding, the \textit{expert} agent reasons over the retrieved code fragments, examining control flow and data constraints surrounding each vulnerable API invocation.
In the Kylin project (Listing~\ref{lst:vulnanalysis-kylin}), \tool identifies built-in safeguards, including password-pattern validation checks (Listing~\ref{lst:vulnprompt-kylin}), and correctly determines that the API usage is secure despite the presence of the vulnerable call. Conversely, in the Nacos project (Listing~\ref{lst:vulnanalysis-nacos}), the expert agent’s \codefont{poc\_understanding} field captures a feasible trigger condition, its resulting impact (e.g., throwing an \codefont{IllegalArgumentException} when a null password is provided), and the exploitation path.
\ying{Without PoV guidance, \tool lacks vulnerability-specific semantics to validate exploitation conditions and paths, causing it to classify both projects as vulnerable based solely on vulnerable API usage.}

\begin{figure}[hbtp]
\begin{minipage}[t]{0.47\linewidth}
\textbf{(a)} Nacos (vulnerable)
\begin{MintedVerbatim}[commandchars=\\\{\},codes={\catcode`\$=3\catcode`\^=7\catcode`\_=8\relax},fontsize=\tiny,numberblanklines=true,breaklines=true,frame=lines,breakautoindent=true,breaksymbolindentleftnchars=4]
\PYG{n+nt}{\PYGZdq{}is\PYGZus{}vulnerable\PYGZdq{}}\PYG{p}{:}\PYG{+w}{ }\PYG{k+kc}{true}\PYG{p}{,}
\PYG{n+nt}{\PYGZdq{}analysis\PYGZdq{}}\PYG{p}{:}
\PYG{+w}{  }\PYG{n+nt}{\PYGZdq{}pov\PYGZus{}understanding\PYGZdq{}}\PYG{p}{:}
\PYG{+w}{    }\PYG{n+nt}{\PYGZdq{}category\PYGZdq{}}\PYG{p}{:}\PYG{+w}{ }\PYG{l+s+s2}{\PYGZdq{}Special Trigger Flow\PYGZdq{}}\PYG{p}{,}
\PYG{+w}{      }\PYG{n+nt}{\PYGZdq{}attack\PYGZus{}mechanism\PYGZdq{}}\PYG{p}{:}
\PYG{+w}{        }\PYG{n+nt}{\PYGZdq{}vulnerable\PYGZus{}api\PYGZdq{}}\PYG{p}{:}\PYG{+w}{ }\PYG{l+s+s2}{\PYGZdq{}BCryptPasswordEncoder.encode\PYGZdq{}}\PYG{p}{,}
\PYG{+w}{        }\PYG{n+nt}{\PYGZdq{}required\PYGZus{}conditions\PYGZdq{}}\PYG{p}{:}
\PYG{+w}{        }\PYG{n+nt}{\PYGZdq{}condition\PYGZdq{}}\PYG{p}{:}\PYG{+w}{ }\PYG{l+s+s2}{\PYGZdq{}raw password is null\PYGZdq{}}\PYG{p}{,}
\PYG{+w}{        }\PYG{n+nt}{\PYGZdq{}explanation\PYGZdq{}}\PYG{p}{:}\PYG{+w}{ }\PYG{l+s+s2}{\PYGZdq{}The encode method of BCryptPasswordEncoder throws an IllegalArgumentException when the raw password is null.\PYGZdq{}}\PYG{p}{,}
\PYG{+w}{        }\PYG{n+nt}{\PYGZdq{}attack\PYGZus{}flow\PYGZdq{}}\PYG{p}{:}\PYG{+w}{ }\PYG{l+s+s2}{\PYGZdq{}When a null value is passed to the encode method, it triggers an IllegalArgumentException.\PYGZdq{}}\PYG{p}{,}
\PYG{+w}{        }\PYG{n+nt}{\PYGZdq{}impact\PYGZdq{}}\PYG{p}{:}\PYG{+w}{ }\PYG{n+nt}{\PYGZdq{}Potential denial of service if the application does not handle the exception properly.\PYGZdq{}}
\PYG{+w}{  }\PYG{n+nt}{\PYGZdq{}code\PYGZus{}analysis\PYGZdq{}}\PYG{p}{:}
\PYG{+w}{    }\PYG{n+nt}{\PYGZdq{}pattern\PYGZus{}match\PYGZdq{}}\PYG{p}{:}
\PYG{+w}{    }\PYG{n+nt}{\PYGZdq{}matches\PYGZdq{}}\PYG{p}{:}\PYG{+w}{ }\PYG{k+kc}{true}\PYG{p}{,}
\PYG{+w}{    }\PYG{n+nt}{\PYGZdq{}explanation\PYGZdq{}}\PYG{p}{:}\PYG{+w}{ }\PYG{n+nt}{\PYGZdq{}The code uses BCryptPasswordEncoder.encode, which is the vulnerable API identified in the POC.\PYGZdq{}}
\PYG{+w}{  }\PYG{esc}{...}
\PYG{+w}{  }\PYG{n+nt}{\PYGZdq{}conclusion\PYGZdq{}}\PYG{p}{:}
\PYG{+w}{    }\PYG{n+nt}{\PYGZdq{}verdict\PYGZdq{}}\PYG{p}{:}\PYG{+w}{ }\PYG{k+kc}{true}\PYG{p}{,}
\PYG{+w}{    }\PYG{n+nt}{\PYGZdq{}explanation\PYGZdq{}}\PYG{p}{:}\PYG{+w}{ }\PYG{l+s+s2}{\PYGZdq{}The code is vulnerable because it directly uses the encode method without checking for null inputs, which can lead to an IllegalArgumentException.\PYGZdq{}}\PYG{p}{,}
\end{MintedVerbatim}
\end{minipage}
\hfill
\begin{minipage}[t]{0.50\linewidth}
\textbf{(b)} Kylin (secure)
\begin{MintedVerbatim}[commandchars=\\\{\},codes={\catcode`\$=3\catcode`\^=7\catcode`\_=8\relax},fontsize=\tiny,numberblanklines=true,breaklines=true,frame=lines,breakautoindent=true,breaksymbolindentleftnchars=4]
\PYG{n+nt}{\PYGZdq{}is\PYGZus{}vulnerable\PYGZdq{}}\PYG{p}{:}\PYG{+w}{ }\PYG{k+kc}{false}\PYG{p}{,}
\PYG{n+nt}{\PYGZdq{}analysis\PYGZdq{}}\PYG{p}{:}
\PYG{esc}{...}
\PYG{+w}{  }\PYG{n+nt}{\PYGZdq{}code\PYGZus{}analysis\PYGZdq{}}\PYG{p}{:}
\PYG{+w}{    }\PYG{n+nt}{\PYGZdq{}pattern\PYGZus{}match\PYGZdq{}}\PYG{p}{:}
\PYG{+w}{    }\PYG{n+nt}{\PYGZdq{}matches\PYGZdq{}}\PYG{p}{:}\PYG{+w}{ }\PYG{k+kc}{true}\PYG{p}{,}
\PYG{+w}{    }\PYG{n+nt}{\PYGZdq{}explanation\PYGZdq{}}\PYG{p}{:}\PYG{+w}{ }\PYG{l+s+s2}{\PYGZdq{}The code uses BCryptPasswordEncoder.encode() in pwdEncode() method\PYGZdq{}}\PYG{p}{,}
\PYG{+w}{    }\PYG{n+nt}{\PYGZdq{}mitigations\PYGZdq{}}\PYG{p}{:}\PYG{+w}{ }\PYG{p}{[}
\PYG{+w}{      }\PYG{n+nt}{\PYGZdq{}type\PYGZdq{}}\PYG{p}{:}\PYG{+w}{ }\PYG{l+s+s2}{\PYGZdq{}Input Validation\PYGZdq{}}\PYG{p}{,}
\PYG{+w}{      }\PYG{n+nt}{\PYGZdq{}effectiveness\PYGZdq{}}\PYG{p}{:}\PYG{+w}{ }\PYG{l+s+s2}{\PYGZdq{}Strong \PYGZhy{} Multiple layers of validation prevent null passwords\PYGZdq{}}
\PYG{+w}{      }\PYG{p}{,}
\PYG{+w}{      }\PYG{n+nt}{\PYGZdq{}type\PYGZdq{}}\PYG{p}{:}\PYG{+w}{ }\PYG{l+s+s2}{\PYGZdq{}Pattern Matching\PYGZdq{}}\PYG{p}{,}
\PYG{+w}{      }\PYG{n+nt}{\PYGZdq{}effectiveness\PYGZdq{}}\PYG{p}{:}\PYG{+w}{ }\PYG{l+s+s2}{\PYGZdq{}Strong \PYGZhy{} Passwords must match specific pattern\PYGZdq{}}
\PYG{+w}{      }\PYG{p}{]}
\PYG{+w}{  }\PYG{n+nt}{\PYGZdq{}conclusion\PYGZdq{}}\PYG{p}{:}
\PYG{+w}{    }\PYG{n+nt}{\PYGZdq{}verdict\PYGZdq{}}\PYG{p}{:}\PYG{+w}{ }\PYG{k+kc}{false}\PYG{p}{,}
\PYG{+w}{    }\PYG{n+nt}{\PYGZdq{}explanation\PYGZdq{}}\PYG{p}{:}\PYG{+w}{ }\PYG{l+s+s2}{\PYGZdq{}The code has multiple safeguards preventing null passwords from reaching the encoder:\PYGZbs{}n1. Password pattern validation (checkNewPwdRule)\PYGZbs{}n2. BCrypt pattern checking\PYGZbs{}n3. Input validation in the controller\PYGZdq{}}
\end{MintedVerbatim}
\end{minipage}
\captionof{listing}{\tool's analysis: (a)~insecure usage in Nacos~\cite{nacos} — exploit triggered by null password; (b)~secure usage in Kylin~\cite{apache_kylin_user_controller} — mitigated by input validation.}
\label{lst:vulnanalysis-nacos}\label{lst:vulnanalysis-kylin}
\end{figure}

\begin{adjustwidth} {0mm}{0mm}
\begin{tcolorbox}
[colback=gray!8, size=title, boxsep=1mm, colframe=black, before={\vskip1mm}, after={\vskip0mm}]
\textbf{Finding 3:}
\emph{\ying{Each component plays a critical role in \tool's effectiveness. AST-based indexing and RAG enable retrieval of security-relevant inter-procedural context, whereas PoV-guided semantic reasoning allows \tool to understand vulnerability-specific trigger conditions, impacts, and exploitation paths.}}
\end{tcolorbox}
\end{adjustwidth}

\subsection{RQ3 (Baseline Comparison)}\label{sec:rq3}

To assess the performance of \tool relative to existing solutions, we compare it against naive LLM and four baseline tools 
from two categories: static programming analysis tools and deep learning-based software approaches.
\begin{itemize}
    \item \textbf{Naive LLM} is our baseline, where we use GPT-4o and provide the entire codebase with a system prompt that includes the PoV test.
    \item \textbf{Codex}~\cite{openai_codex_cli} is an OpenAI LLM agent that autonomously explores and analyzes repositories.
    \item \textbf{VAScanner}~\cite{vascanner-github} uses call-graph analysis to determine whether vulnerable APIs from library dependencies are reachable and actually invoked within the application code.
    \item \textbf{Eclipse Steady}~\cite{ponta2020detection} performs reachability analysis by combining static and dynamic techniques, tracing execution paths from known vulnerable APIs to their invocation points in the application. It then produces detailed assessments of vulnerability exploitation.
    \item \textbf{LineVul}~\cite{fu2022linevul} is a transformer-based model leveraging CodeBERT~\cite{feng2020codebert} for code representation and vulnerabilities at line level. 
    \item \textbf{DeepDFA}~\cite{steenhoek2024dataflow} combines dataflow analysis (DFA) with graph neural networks (GNNs) for detecting software vulnerabilities. It encodes dataflow information at each node of a program's control flow graph and learns data usage patterns. 
\end{itemize}

\textbf{Experiment settings.} To ensure a fair comparison, we run experiments for all tools on the same machine.
Naive LLM uses GPT-4o with a 128K context window, packing each codebase's source code with Repomix 1.6.0~\cite{yamada2024repomix}. Codex uses the gpt-5.2 model in full-auto mode, with the PoV test and vulnerable API signature as the task prompt; we disable web search to ensure the agent relies solely on repository exploration. VAScanner runs on its published code~\cite{vascanner-github} with default settings, with program-issue fixes we applied, while Eclipse Steady version 3.2.5 runs with its default configuration; we set a 150-minute per-project analysis time limit for both tools. For the machine-learning-based tools LineVul and DeepDFA, we use their default parameter settings with function-level inputs. \lx{Since neither tool was designed for project-level exploitability assessment, we include them as reference.} For \tool, we use the same settings as described in Section~\ref{sec:implementation}.

\begin{table}[t!]
\caption{Runtime performance of five tools}
\scriptsize
\centering
\begin{tabular}{|l|c|c|c|}
\hline
\textbf{Tools} & \textbf{ Time (Total)} & \textbf{Time (Average)} & \textbf{Failed Cases} \\
\hline
\textbf{\tool} & 275 mins & 5 mins & 0 \\
\hline
\textbf{Naive LLM} & 3 mins & 9 secs & 38 \\
\hline
\textbf{Codex} & 122 mins & 128 secs & 0 \\
\hline
\textbf{Eclipse Steady} & 722 mins & 13 mins & 18 \\
\hline
\textbf{VAScanner} & 41 mins & 2 mins  & 35 \\
\hline
\textbf{DeepDFA} & 364 mins & 6 mins & 0 \\
\hline
\textbf{LineVul} & 361 mins & 6 mins & 0 \\
\hline

\end{tabular}
\label{tab:runtime-performance}
\end{table}

CognixShield achieves an overall F1 of 89\% and A of 84\%, outperforming Naive LLM (F1: 42\%, A: 30\%), Codex (F1: 76\%, A: 65\%), LineVul (F1: 42\%, A: 42\%), EclipseSteady (F1: 40\%, A: 28\%), VAScanner (F1: 33\%, A: 25\%), and DeepDFA (F1: 33\%, A: 24\%).
The naive LLM achieved only 28\% recall across 57 projects; due to the agentic RAG design, \tool uses 5.7× fewer LLM tokens at significantly lower cost.

Although Codex outperforms all other baselines, \tool is still ahead by 13\% in F1 and 19\% in accuracy. Codex's iterative grep-and-read strategy struggles with cross-file reachability: when a vulnerable sink requires traversing multi-file chains (AOP aspects, factories, runtime configuration), it commits to a verdict before resolving the full chain, whereas \tool's RAG retrieval co-locates all relevant fragments in a single prompt. For example, in \texttt{PLMCodeTemplate} (CVE-2020-26217), the agent found no live callers of \codefont{XMLConfig.toBean()} and concluded the sink was dead code. The sink is in fact reachable via a four-file Spring AOP chain: a \codefont{@Log}-annotated controller triggers \codefont{LogAOP}, which evaluates attacker-controlled SPEL via \codefont{SPELUtil}, enabling \codefont{T(XMLConfig).toBean(...)} as a gadget. The agent read \codefont{XMLConfig.java} but never traversed the annotation-to-aspect-to-SPEL path, while \tool's retrieval co-located all four files in a single context, correctly identifying the project as vulnerable.



VAScanner achieves precision of 64\%, but yields low recall (23\%) due to scalability failures on large codebases. Eclipse Steady exhibits a similar trend, achieving 30\% R and an F1 of 40\%, failing to analyze 18 projects due to limitations in building call graphs for complex projects. 
The DL-based tools, LineVul and DeepDFA, achieve low recall rates of 30\% and 15\%, respectively. These approaches overlook a substantial number of vulnerable API usages. This limitation may be due to their function-level detection strategy, where each function is analyzed in isolation. Detecting vulnerable API usage, however, often requires a broader context that spans across multiple functions and files.
\textit{Runtime Performance.} As shown in Table~\ref{tab:runtime-performance}, \tool averages 5 minutes per project (275 minutes total) and completed all 57 applications without failures. Eclipse Steady required 722 minutes and failed on 18 projects; VAScanner finished in 41 minutes but failed on 35 cases. DL-based tools averaged 6 minutes per project including training time. \tool eliminates training requirements and achieves 100\% project coverage, making it more practical for real-world deployment.

\begin{adjustwidth} {0mm}{0mm}
\begin{tcolorbox}
[colback=gray!8, size=title, boxsep=1mm, colframe=black, before={\vskip1mm}, after={\vskip0mm}]
\textbf{Finding 4:}
\emph{\newx{CognixShield outperforms all baselines across every metric (F1 89\%, accuracy 84\%). The strongest baseline, the Codex agent, reaches 76\% F1 and 65\% accuracy, while the remaining static-analysis, DL-based, and naive-LLM baselines fall below 42\% accuracy.} It demonstrates better scalability and efficiency on large real-world projects, overcoming memory and dependency challenges faced by existing approaches.}
\end{tcolorbox}
\end{adjustwidth}
\vspace{-0.5em}
\subsection{RQ4 (Robustness Analysis)}
\label{sec:rq4}
\begin{figure*}[hbtp]
    \centering
    \includegraphics[width=\textwidth]{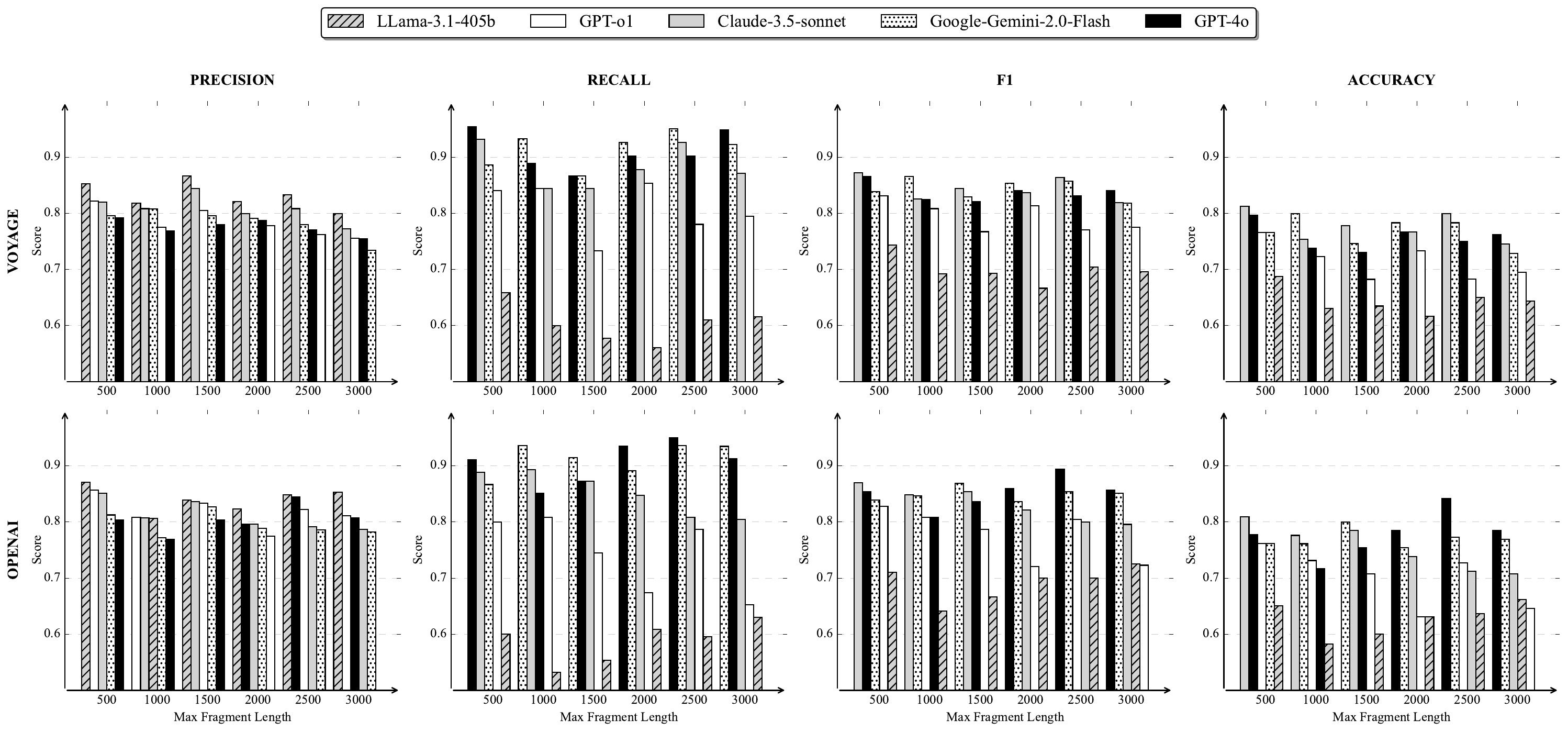}  
    \caption{Performance evaluation of LLM-embedding model combinations across six code fragment lengths. The top legend denotes the LLMs, while the left legend indicates the embedding models. Each subplot corresponds to a distinct evaluation metric as listed (Precision, Recall, F1-score, and Accuracy).
    }
    \label{fig:llm-combined-comparison-vertical}
\end{figure*}

\textbf{Experiment settings:} we evaluate \tool on two popular text embedding models \emph{voyage-code-3}\cite{voyage_code_3}, and \emph{OpenAI text-embedding-3-small}~\cite{openai_embeddings}; five state-of-the-art closed-source LLMs and one open-source LLM: \fo~\cite{openai_gpt4o}, \gpto~\cite{openai_gpt4o}, \claud~\cite{anthropic_claude_sonnet}, \gemini~\cite{gemini} and \fb~\cite{meta_llama_3_1}, while varying the maximum code fragment size ($\theta$) over 500, 1000, 1500, 2000, 2500, and 3000. Figure \ref{fig:llm-combined-comparison-vertical} presents the performance results across four key metrics (P, R, F1, A). The first row shows results using the \emph{voyage-code-3} embedding, while the second row shows results using the \emph{OpenAI text-embedding-3-small} embedding.

\textbf{Embedding Models:} \tool shows minimal sensitivity to the choice of embedding models. The \textit{openAI} embedding achieves more stable precision and the highest F1 of 90\% (P: 84\%, R: 95\%) with \fo at $\theta = 2500$ tokens, while the \textit{voyage} embedding consistently yields higher recall (up to 96\% for \fo) but lower precision, with recall differences reaching up to 12\% between the two embeddings.

\textbf{Impact of LLMs:} Across five different LLMs, \tool achieves precision ranging from 74\% to 87\% and recall from 53\% to 96\%. \fo achieves the highest overall performance (F1: 89\%, A: 84\%), while \fb reaches the highest precision (87\%) but the lowest recall (53\%--66\%), resulting in the worst F1 (below 72.5\%) and accuracy (below 66\%). These findings indicate that the choice of LLM substantially affects detection effectiveness, with \fo demonstrating the strongest capability.




\textbf{Code Fragmentation Length:} Most models maintain stable precision across fragment sizes, while recall improves for \fo and \gemini as fragment length increases (e.g., \fo's recall grows from 91\% at 500 tokens to 95\% at 2500 tokens). Beyond 2500 tokens, some models show degradation: \gpto's F1 drops from 83\% to 72\% when increasing from 500 to 3000 tokens, likely due to overly long fragments introducing noise. This suggests that 2000–2500 tokens represents the optimal balance for large-scale vulnerability detection.

\begin{adjustwidth} {0mm}{0mm}
\begin{tcolorbox}
[colback=gray!8, size=title, boxsep=1mm, colframe=black, before={\vskip1mm}, after={\vskip0mm}]
\textbf{Finding 5:}
\emph{\tool demonstrates stable performance across embedding models and fragment sizes, while LLM choice has a substantial impact on detection effectiveness (F1: 64\%–89\%, A: 58\%–84\%). \fo consistently achieves the best results (F1: 81\%–89\%, A: 75\%–84\%), whereas \fb underperforms (F1: 64\%–73\%). Code segments of 2,000–2,500 tokens with \textit{openAI} embeddings yield the highest overall performance.}
\end{tcolorbox}
\end{adjustwidth}



\section{Threats to validity}
\label{sec:threats}

\textit{Internal Validity.} 1) Parameter Sensitivity: Our approach's performance may be sensitive to parameters like similarity thresholds and LLM settings. While we set the temperature to 0 for reproducibility, other parameters (e.g., similarity threshold) may still influence results.
2) Ground Truth Accuracy: The manual labeling data requires significant effort. To ensure labeling accuracy, two authors who have industry experience in security domains labeled the data. However, this approach might impact the overall accuracy of the classification. 3) Prompt Template Limitation: We used fixed prompt templates for the LLM. This could limit the range of vulnerabilities detected and affect the accuracy of our results. 4) Implementation Choices: \tool's effectiveness may be influenced by specific implementation choices (e.g., AST parser, encoder models).


\textit{External Validity.} The generalizability of our findings may be limited by the following factors:
1) The principle of our retrieval-augmented analysis is language-agnostic,
  but our concrete implementation includes Java-specific components, so
  the generalizability of our quantitative findings to other languages is
  not directly established. We chose to instantiate the approach on Java,
  motivated by Java's prevalence in open-source software and growing
  security concerns over supply chain attacks in popular Java libraries.
  2) Our observations and conclusions are based on the third-party
  datasets~\cite{kang2022test} complemented by additional data collected
  from GitHub. While these datasets offer valuable insights, they do not
  cover all possible scenarios or vulnerabilities in real-world
  applications.
  3) Our study relies on exemplar security tests from libraries to
  illustrate vulnerability contexts. According to~\cite{kang2022test},
  over 51\% of vulnerabilities have associated test cases. While this
  approach is effective for evaluation, it may be limited by the
  availability of PoV tests.


\section{Related Work}

\textbf{Vulnerable API Detection} 
Researchers and engineers created tools to detect the invocation of vulnerable APIs or misuse of APIs~\cite{singleton2020cryptotutor,ponta2020detection, Zhang2022,fischer2017stack}. These tools are designed for identifying insecure cryptographic API usage patterns that may introduce security risk. Specifically, tools such as  CryptoTutor~\cite{singleton2020cryptotutor}, and SEADER~\cite{Zhang2022} adopted static analysis to identify whether Java cryptographic APIs are invoked with secure parameter configurations and correct sequential orders. 
Fischer et al.~\cite{fischer2017stack}  trained models on labeled code snippets that demonstrate both secure and insecure cryptographic API usage. These models are then applied to detect potentially insecure API usage patterns in new Java code snippets, learning from examples to identify code that deviates from secure usage patterns.
\newx{For vulnerable APIs introduced through library dependencies, version-based scanners such as DependencyCheck~\cite{dependency-check} match declared dependency versions against vulnerability databases, whereas reachability-based tools decide whether the vulnerable API is actually invoked: VAScanner~\cite{zhang2024does} and Eclipse Steady~\cite{ponta2020detection} build call graphs from the application to the vulnerable API, and ViperScan~\cite{viperscan} further uses an LLM to distill exploit conditions from security patches into static reachability rules. These tools rely on syntactic patterns or reachability, and do not reason about whether a reached API is actually exploitable in the surrounding program context.}

\textbf{Deep Learning based Software Vulnerability Detection}
Deep learning-based software vulnerability detector approaches~\cite{fu2022linevul, ivdetect, li2018vuldeepecker, steenhoek2024dataflow, cao2024coca, sysevr} have been proposed to automatically learn vulnerability patterns from various code representations. IVDetect~\cite{ivdetect} utilized RNN-based models to generate code representations from source code and identify vulnerable code patterns. LineVul~\cite{fu2022linevul} leverages CodeBERT, a pre-trained model for programming language, to generate code vector representations and employs BERT's self-attention layers to capture long-term dependencies in code sequences. DeepDFA~\cite{steenhoek2024dataflow} encodes dataflow analysis information into graph neural networks to better capture program semantics for vulnerability detection.  These methods often focus on detecting vulnerabilities within individual functions or code snippets rather than considering the broader inter-procedural context required for dependency vulnerability detection.



\noindent\textbf{LLM for Vulnerability Detection}
Recent work has explored LLMs' capabilities in security tasks~\cite{secllmholmes,  zhou2024large,liu2024generating}. SecLLMHolmes~\cite{secllmholmes} evaluates the ability of various LLMs to detect security-related bugs under different configurations, concluding that current models are non-deterministic and not robust. 
Liu et al.~\cite{liu2024generating} introduce GPTAid which adopted LLM to infer API misuse parameter rules and adopts existing tools (CodeQL~\cite{codeql}) for detection.
Several studies investigated vulnerable code repair~\cite{examining-zero-shot, on-fine-tuning,  automated-program-repair} through LLMs. For instance, Pearce et al.~\cite{examining-zero-shot} examine LLMs in repairing vulnerable code in a zero-shot prompt, finding that while models can successfully repair hand-crafted examples, they struggle with complex, real-world cases.
\newx{To move beyond single functions, Vul-RAG~\cite{vulrag} retrieves distilled vulnerability knowledge to judge individual functions, and LLMxCPG~\cite{llmxcpg} uses code-property-graph slices to steer the LLM toward vulnerability-relevant code, but both remain function- or slice-level rather than repository-scale. \tool instead combines AST fragmentation, embedding retrieval, agentic RAG, and LLM reasoning to assemble repository-scale inter-procedural context and determine exploitability.}

\section{Conclusion}

We presented \tool, a novel Agentic RAG approach for detecting vulnerable API usage in large-scale codebases, which combines inter-procedural context retrieval with LLM-driven vulnerability code semantic reasoning. Unlike traditional analyzers, \tool constructs repo-level program contexts capturing control flow and data constraints, enabling accurate distinction between exploitable and safe API usages.
This work demonstrates that LLM-based techniques, when integrated with domain knowledge, can make vulnerability detection more practical and scalable.
Future directions include extending \tool to support additional languages, such as Golang and JavaScript, developing adaptive context retrieval strategies that adjust to code complexity, and enhancing interpretability by providing actionable fix suggestions alongside detection results.

\section*{Acknowledgement}
This work has been partially supported by the Wake Forest Pilot Research Grant. We thank the anonymous reviewers for their valuable comments on an earlier version of this paper.

\bibliography{reference}


\end{document}